\documentclass[12pt]{article}
\input epsf
\newcommand{\sect}[1]{\section{#1}\setcounter{equation}{0}}

\newcommand{\OL}[1]{ \hspace{1pt}\overline{\hspace{-2.5pt}#1
  \hspace{-1.2pt}}\hspace{2pt} }
\newcommand{\skyp}[1]{}

\begin{document}

\bigskip
\hskip 5in\vbox{\baselineskip12pt \hbox{%NSF-ITP-03-xxx
} }
\bigskip\bigskip

\centerline{\Large AdS Holography in the Penrose Limit}
\bigskip
\bigskip
\bigskip
\centerline{\bf Nelia Mann}
\medskip
\centerline{Department of Physics} \centerline{University of
California} \centerline{Santa Barbara, CA\ \ 93106}
\centerline{\it nelia@physics.ucsb.edu}
\bigskip
\centerline{\bf Joseph Polchinski}
\medskip
\centerline{Kavli Institute for Theoretical Physics}
\centerline{University of California} \centerline{Santa Barbara,
CA\ \ 93106-4030} \centerline{\it joep@kitp.ucsb.edu}
\bigskip
\bigskip

\begin{abstract}
We study the holographic description of string theory in a plane wave
spacetime by taking the Penrose limit of the usual AdS/CFT
correspondence.  We consider three-point functions with two BMN operators
and one non-BMN operator; the latter should go over to a perturbation
of the dual CFT.  On the string side we take the Penrose limit of the
metric perturbation produced by the non-BMN operator, and the BMN state
propagates in this perturbed background.  The work of Lee, Minwalla,
Rangamani, and Seiberg shows that for chiral operators the AdS
three-point functions agree with those calculated in the free gauge
theory.  However, when this is reduced to an effective plane wave
amplitude by truncating the amplitude to propagate from the AdS boundary
to the Penrose geodesic, we find a puzzling mismatch.  We discuss possible
resolutions, and future directions.
\end{abstract}

\newpage
\baselineskip=17.6pt

\section{Introduction}

Recently, the plane wave (Penrose) limit of the AdS/CFT
correspondence has received much attention~\cite{BMN}.  In the
field theory, this amounts to selecting a single generator $J$ of
the $SO(6)$ $R$-symmetry, and considering only operators that have
a large charge (and small $\Delta - J$) with respect to this
symmetry.  On the supergravity side of the correspondence, one
takes a limit that focuses in on a single null geodesic, yielding
the plane wave metric~\cite{BFHP}.  There exists an exact
light-cone quantization of string theory in this
background~\cite{metplane} and a dictionary relating the string
states to the CFT operators.  Furthermore, some quantities can be
expanded in terms of constants which are simultaneously small on
both sides of the correspondence~\cite{BMN,KPSS,GMR,camb}. Plane
wave physics is of particular interest because of the transparent
nature of the string/field theory duality; it yields a dictionary
between objects which have manifestly similar structures.  In
particular, when analyzing the spectrum, one can see in the field
theory where the ``string" arises.

In the full AdS/CFT duality, there remain important problems of
principle. One is the gauge theory origin of the approximate
locality in the bulk spacetime.  Another, related to this, is a
more complete understanding of the quantum mechanics of black
holes (for recent discussion see ref.~\cite{KOS}). A third is the
extension of the duality to more general spacetimes, in particular
those with interesting cosmologies.  One might hope that the more
transparent duality of the plane wave limit would provide some
ideas in these directions, but this has not yet been realized.
Indeed, there seems to be an important ingredient missing from the
plane wave picture.  Currently string theory in the plane wave
background is understood as being dual to a limit of the gauge
theory, but it is widely anticipated that there is some smaller
CFT, not yet identified, which is precisely dual to the string
side.

Our goal in this paper is to obtain some insight into this
question by looking at holographic observables that are analogous
to those studied in the full AdS/CFT correspondence~\cite{GKPW}.
These would correspond to local operators in the dual CFT, and
would map to perturbations of the boundary conditions in the plane
wave spacetime. An understanding of the properties of these local
operators may give some insight into the nature of the CFT itself.

There has been some general discussion of holography in this
context, focussing on the geometry of the plane wave
spacetime~\cite{ppholo}, but no complete understanding has
emerged.  We will approach the question in a slightly different
way, by starting with a CFT matrix element and its dual in the
$AdS_5 \times S^5$ theory, where the holographic dictionary is
known, and taking the Penrose limit.  It is natural to hold fixed
the dimension and quantum numbers of the Hamiltonian perturbation
as we take the limit, so the simplest matrix element to consider
would be two BMN (large-$J$) operators, to prepare the initial and
final string states, and one non-BMN operator corresponding to the
perturbation.

The inclusion of a non-BMN operator is consistent with the
arguments of refs.~\cite{Vbit,BKPSS} that in order to take the
plane wave limit, all BMN operators must be taken to $t = \pm
\infty$, and so correspond to incoming and outgoing strings rather
than perturbations of the Hamiltonian. Thus the non-BMN operators
cannot naturally be regarded as limits of BMN operators, but have
a different interpretation. Most of the previous work on
three-point functions, beginning with refs.~\cite{3bmn}, has
focused on three BMN operators, corresponding to a three-string
interaction.

We begin with the three-point AdS/CFT calculation of  Lee,
Minwalla, Rangamani, and Seiberg (LMRS)~\cite{LMRS} and adapt it
to the plane wave limit.  That is, we start with the LMRS result
for the metric perturbation corresponding to the insertion of a
chiral operator, and take its limit.  We then calculate the
amplitude for the string to propagate through this perturbed
background.\footnote {We work to leading order in the effective
string coupling $J^2/N$, so we are considering only planar
world-sheets.} For a three-point function the position dependence
is determined entirely by conformal invariance, so the
normalization is the only nontrivial aspect, and we focus on this.

Of course, the results of LMRS already show that for three chiral
operators the duality works, in that the normalization in the free
gauge theory exactly matches that in the AdS calculation, and we
reproduce this agreement in our framework.  However, we argue that
the agreement actually presents a puzzle.  Local operators in the
dual to the plane wave geometry should differ from those in the
full $AdS_5 \times S^5$, in that the propagation from the $AdS_5
\times S^5$ boundary to the limiting geodesic is not part of the
plane wave theory and this propagation amplitude should be
truncated.  When this is done the form of the amplitude changes so
that it no longer has an obvious match to the dual theory.  This
may be related to problems found by Spradlin and Volovich in the
corresponding four-point amplitude (unpublished).

In section two we discuss the field theory amplitude.  We consider
the three point function with one BMN operator mapped to negative
infinity in global time and the other BMN operator to the positive
infinity, thus acting as creation and destruction operators for
the string state; the non-BMN operator is fixed at finite time.
In section three we discuss the string amplitude.  We first
calculate the metric perturbation that corresponds to the non-BMN
operator via the AdS/CFT dictionary and take its Penrose limit.
We then look at the light-cone quantized string action in this
perturbed background, and calculate the amplitude for the initial
string to propagate through the perturbed background to the given
final state.  Our main calculational result is to reproduce the
result of LMRS for three chiral operators, but in a framework that
readily extends to nonchiral BMN operators and to amplitudes with
additional non-BMN operators. In the final section we discuss the
discrepancy that we have found, and suggest several possible
resolutions.  We also discuss future directions.

\section{Field theory calculation}

We first review the calculation of LMRS\mbox{}. In the $D = 4$,
$\mathcal{N} = 4$, $SU(N)$ supersymmetric Yang-Mills theory, the
operators that we are studying will all be of the form
\begin{equation}
O^I = \kappa\mathcal{C}^I_{i_1 ... i_k} {\rm
Tr}(\phi^{i_1}\phi^{i_2} \cdots \phi^{i_k}) \equiv \kappa
\mathcal{C}^I(\phi)\ .
\end{equation}
The six scalars $\phi^i$ form the fundamental representation of
the $SO(6)$ $R$-symmetry of the field theory. Here $\mathcal{C}^I$
is a totally symmetric traceless rank $k$ tensor of $SO(6)$,
normalized such that
\begin{equation}
\langle \mathcal{C}^{I_1} \mathcal{C}^{I_2 *} \rangle \equiv
\mathcal{C}^{I_1}_{i_1 ... i_k} \mathcal{C}^{I_2*}_{ i_1 ... i_k}
= \delta^{I_1 I_2}\ .
\end{equation}
The constant $\kappa = (2\pi)^k /(g_{\rm YM}^2 N)^{k/2}\sqrt{k} $
is fixed to normalize the two-point function,
\begin{equation}
\langle O^{I_1}(x_1)O^{I_2 *}(x_2) \rangle =
\frac{\delta^{I_1I_2}}{|x_{12}|^{2k}}\ . \label{twopoint}
\end{equation}
The three-point function is then
\begin{equation}
\langle O^{I_1}(x_1)O^{I_2}(x_2)O^{I_3}(x_3) \rangle =
\frac{1}{N}\frac{\sqrt{k_1k_2k_3}\, \langle
\mathcal{C}^{I_1}\mathcal{C}^{I_2}\mathcal{C}^{I_3}\rangle}
{|x_{12}|^{2\alpha_3}|x_{23}|^{2\alpha_1}|x_{13}|^{2\alpha_2}}
\end{equation}
  where $k_1, k_2, k_3$ are the ranks of
$\mathcal{C}^{I_2}, \mathcal{C}^{I_2}, \mathcal{C}^{I_3}$.  The
$\alpha_i$ are defined cyclically such that $\alpha_1 = (k_2 + k_3
- k_1)/{2}$; $\alpha_1$ is the number of contractions between
$O^{I_2}$ and $O^{I_3}$, and so on.  The position-dependence of
the two- and three-point functions is determined by conformal
symmetry.

To take the BMN limit, we decompose $SO(6) \to  SO(2) \times
SO(4)$.  The combinations $Z = (\phi_5 + i\phi_6)\sqrt 2$ and $\OL
Z = (\phi_5 - i\phi_6)/\sqrt 2$ are eigenvectors of the $SO(2)$
generator. The three operators that we will consider are
\begin{eqnarray}
O^{I_1} &=& \tilde\kappa_1\tilde{\mathcal{C}}^{I_1}_{i_1 ...
i_{\tilde{k}_1}} {\rm Tr}(Z^J\phi^{i_1} \cdots
\phi^{i_{\tilde{k}_1}} + \mbox{permutations})\ ,\quad {k_1 \choose
J }\ \mbox{terms}
\ , \nonumber\\
O^{I_2} &=& \tilde\kappa_2\tilde{\mathcal{C}}^{I_2}_{i_1 ...
i_{\tilde{k}_2}}{\rm Tr}(\OL{Z}^J\phi^{i_1} \cdots
\phi^{i_{\tilde{k}_2}} + \mbox{permutations})\ ,\quad
{k_2 \choose J }\ \mbox{terms} \ ,\nonumber\\
O^{I_3} &=& \tilde\kappa_3\tilde{\mathcal{C}}^{I_3}_{i_1 ...
i_{\tilde{k}_3}}{\rm Tr}(\phi^{i_1} \cdots
\phi^{i_{\tilde{k}_3}})\ .
\end{eqnarray}
The $\tilde\mathcal{C}^{I}$ are totally symmetric traceless
$SO(4)$ tensors of rank $\tilde{k}$, normalized $\langle
\mathcal{C}^{I_1} \mathcal{C}^{I_2 *}\rangle = \delta^{I_1 I_2}$.
We have defined
\begin{equation}
\tilde k_1 = k_1 - J\ ,\quad \tilde k_2 = k_2 - J\ ,\quad \tilde
k_3 = k_3\ .
\end{equation}
Relating the normalizations of the $SO(6)$ and $SO(4)$ tensors
gives
\begin{equation}
\tilde\kappa_1 = \kappa_1 {k_1 \choose J}^{-1/2}\ ,\quad
\tilde\kappa_2 = \kappa_2 {k_2 \choose J}^{-1/2}\ ,\quad
\tilde\kappa_3 = \kappa_3 \ .
\end{equation}
Define also $\tilde{\alpha}_1 = (\tilde k_2 + \tilde k_3 - \tilde
k_1)/{2}$ and permutations of this.

In the three-point function there are $\alpha_3 \choose J$
permutations that give nonzero contractions between the $Z$
fields, and so
\begin{equation}
\langle \mathcal{C}^{I_1}\mathcal{C}^{I_2}\mathcal{C}^{I_3}\rangle
= {\alpha_3 \choose J} {k_1 \choose J}^{-1/2} {k_2 \choose
J}^{-1/2} \langle \tilde\mathcal{C}^{I_1}\tilde\mathcal{C}^{I_2}
\tilde\mathcal{C}^{I_3}\rangle\ .  \label{ccc}
\end{equation}
Then
\begin{eqnarray}
&& \langle O^{I_1}(x_1)O^{I_2}(x_2)O^{I_3}(x_3) \rangle = \nonumber\\
&&\qquad\qquad\qquad \frac{1}{N} \frac{(\tilde \alpha_3 + J)!
\sqrt{\tilde k_1! \tilde k_2 ! \tilde k_3}} {\tilde\alpha_3!
\sqrt{(\tilde k_1 + J - 1)! (\tilde k_2 + J - 1)!} } \frac{\langle
\tilde\mathcal{C}^{I_1}\tilde\mathcal{C}^{I_2}
\tilde\mathcal{C}^{I_3}\rangle}
{|x_{12}|^{2(J+\tilde\alpha_3)}|x_{23}|^{2\tilde\alpha_1}
|x_{13}|^{2\tilde\alpha_2}}\ .\qquad
\end{eqnarray}
Taking the large-$J$ limit with the tilded quantities held fixed
gives
\begin{equation}
  \langle O^{I_1}(x_1)O^{I_2}(x_2)O^{I_3}(x_3) \rangle \approx
\frac{J^{1 - \frac12 \tilde k_3}}{N } \frac{\sqrt{\tilde k_1!
\tilde k_2 ! \tilde k_3}}{\tilde\alpha_3!} \frac{\langle
\tilde\mathcal{C}^{I_1}\tilde\mathcal{C}^{I_2}
\tilde\mathcal{C}^{I_3}\rangle} {|x_{12}|^{2(J+\tilde\alpha_3)}
|x_{23}|^{2\tilde\alpha_1}|x_{13}|^{2\tilde\alpha_2}}\ .
\label{largej}
\end{equation}

To facilitate the comparison to string theory we rewrite this
vacuum three-point function as an operator matrix element.  We
first convert from the $R^4$ description of the boundary to $S^3
\times R$ via $x_i = e^{\tau_i} \hat x_i$.  Here $\tau_i$ is
Euclidean global time, and $\hat x_i$ is a point on the unit
$S^3$.  The chiral primaries $O^{I}$ each pick up a conformal
factor, $O^{I} = e^{k\tau}O^{\prime I}$, so that for $-\tau_1 \gg
1$ and $\tau_2 \gg 1$ the position dependent factors in
eq.~(\ref{largej}) become $e^{-k_1(\tau_3 - \tau_1) - k_2 (\tau_2
- \tau_3)}$. Similarly the two-point function becomes
\begin{equation}
\langle O^{\prime I_2}(\tau_2,\hat x_2) O^{\prime I_1
*}(\tau_1,\hat x_1) \rangle = \delta^{I_1I_2}e^{-k(\tau_2 -
\tau_1)}\ ,
\end{equation}
so that $O^{I}(\tau,\hat x) | 0 \rangle \to e^{k \tau} |I\rangle$.
Then
\begin{equation}
\langle I_2 | O^{I_3}(\tau_3, \hat x_3) | I_1\rangle \approx
\frac{J^{1 - \frac12 \tilde k_3}}{N } \frac{\sqrt{\tilde k_1!
\tilde k_2 ! \tilde k_3}}{\tilde\alpha_3!}
 e^{-(k_1 - k_2)\tau_3} \langle
\tilde\mathcal{C}^{I_1}\tilde\mathcal{C}^{I_2}
\tilde\mathcal{C}^{I_3}\rangle \ . \label{finalmat}
\end{equation}
Note that this can be Wick-rotated readily from Euclidean to
Minkowski global time.

Since the field theory amplitude does not involve the 't Hooft
parameter (it has been scaled out), the BMN limit is simply large
$J$ with fixed $\tilde C$ and $J^2/N$.  In the final
form~(\ref{finalmat}), $J$ enters only as a power, with the
exponent depending on $\tilde k_3$. Thus to obtain a finite limit
we must rescale $O^{I_3}$ by $J^{\frac12 \tilde k_3 - 1}$.  Note
that for the mass operator, with $\tilde k_3 =2$, no rescaling is
needed: this can couple to the $SO(4)$ $\phi^i$ without
suppression (aside from the $1/N$ from the normalization of
single-trace operators).  For each increase of $\tilde k_3$ by
two, one additional $SO(4)$ propagator must appear in an adjacent
cyclic position, costing a factor of $J$. The factor $\frac{1}{N}$
is the suppression expected from an interaction of three strings,
the tensor contraction follows from $SO(4)$ invariance, and the
$\tau$-dependence from the Heisenberg equation of motion.

\sect{String theory calculation}

According to the AdS/CFT dictionary~\cite{GKPW}, chiral operators
on the boundary correspond to perturbations of the supergravity
fields in the bulk.  Thus the matrix element~(\ref{finalmat})
represents the amplitude for a single string to make a transition
while propagating through the perturbed background produced by the
non-BMN operator $O^{I_3}$.  In section 3.1 we obtain the form of
the perturbed background, and in section 3.2 we use light-cone
string theory to calculate the transition amplitude.

\subsection{The metric perturbation}

We first review the AdS/CFT dictionary~\cite{GKPW}.  We begin with
coordinates in which the Euclidean $AdS_5$ metric is
\begin{equation}
ds^2 = R^2 \frac{dx_0^2 + dx \cdot dx}{x_0^2}\ . \label{metone}
\end{equation}
Each scalar chiral operator $O^I$ gives rise to a scalar field
$\phi^I$ on $AdS_5$, with the dictionary~\cite{BKLT}
\begin{equation}
O^I(x) = c_I \lim_{x_0 \to 0} x_0^{-\Delta_I} \phi^I(x,x_0)\ .
\end{equation}
That is, the fluctuating part of the field $\phi^I$ has the
normalizable behavior $x_0^{\Delta_I}$; after extracting this
$x_0$-dependence, the boundary limit of the bulk field operator is
the boundary operator.  We take the fields $\phi_I$ to be
normalized canonically,
\begin{equation}
S = \frac{1}{2} \sum_I \int_{AdS_5} d^4 x\, dx_0 \sqrt{-g}\,
(\partial_\mu \phi^I \partial^\mu \phi^I + m_I^2 \phi^{I2}) \ ,
\end{equation}
where $m^2_I = \Delta_I(\Delta_I - 4)/R^2$. We take the operators
$O^I$ to have normalized two-point functions~(\ref{twopoint}); for
the present discussion we take a real basis for the fields and
operators.  These two normalizations determine the coefficient
$c_I$.  The canonical scalar propagator satisfies
\begin{equation}
\lim_{x_0' \to 0} x_0'^{-\Delta} \langle 0| \phi(x,x_0)
\phi(x',x_0') |0 \rangle = A(\Delta) \Biggl[\frac{ x_0}{x_0^2 +
(x-x')^2 } \Biggr]^\Delta\ ,
\end{equation}
where
\begin{equation}
A(\Delta) = \frac{\Delta - 1}{2\pi^2 R^3}\ . \label{aofd}
\end{equation}
(For $AdS_{d+1}$, $A(\Delta) = {\Gamma(\Delta)}/{2 \pi^{d/2}
\Gamma(\Delta + 1 - d/2) R^{d-1}}$.) It follows that $c_I^{-2} =
{A(\Delta_I)}$, and that
\begin{equation}
\langle 0 |\, \phi^I(x,x_0)\, O^I(x') \,| 0 \rangle =
{A(\Delta_I)}^{1/2}\Biggl[\frac{x_0}{x_0^2 + (x-x')^2}
\Biggr]^{\Delta_I}\ . \label{phio}
\end{equation}

To take the Penrose limit it will be convenient to put the $AdS_5
\times S^5$ metric in the coordinates
\begin{equation}
ds^2 = R^2\left[-(1+u^2) dt^2 + d\vec u \cdot d\vec u - \frac{u^2
du^2}{1+u^2}\right] + R^2\left[(1-v^2) d\psi^2 + d\vec v \cdot
d\vec v + \frac{v^2 dv^2}{1-v^2}\right] . \label{tenmet}
\end{equation}
The $AdS_5$ part is related to the earlier metric~(\ref{metone})
by a change of coordinates followed by a Wick rotation.  The
coordinate change is
\begin{equation}
e^{\tau} = (x_0^2 + x^2)^{1/2}\ ,\quad \vec u = \vec x/x_0\ .
\end{equation}
  It is then straightforward to Wick rotate $\tau \to
(1-i\epsilon) it$ on  both sides of the duality.  The duality
dictionary~(\ref{phio}) becomes
\begin{equation}
\langle 0 |{\it T}\, \phi^I(t,\vec u)\, O^I(t',\hat x') \,| 0
\rangle = \frac{{A(\Delta_I)}^{1/2} }{2^{\Delta_I} \Bigl(
\sqrt{1+u^2} \cos [(1-i\epsilon)(t-t')] - \vec u \cdot \hat x'
\Bigr)^{\Delta_I}}\ . \label{mink}
\end{equation}
We have included a factor $e^{i \Delta_I t'}$ from the conformal
transformation of $O^I$ in going from $R^4$ to $R \times S^3$.

LMRS express the metric perturbation associated with a chiral
operator in terms of scalar fields $s^I$ on $AdS_5$.  Following
the conventions in LMRS, these are related to canonically
normalized fields by
\begin{equation}
s^I = \frac{2^{(k-3)/2} (k_I+1) \pi R^{3/2}}{N \sqrt{k_I (k_I-1)}
} \phi^I \ ;  \label{norm}
\end{equation}
note that $k_I = \Delta_I$. The perturbation takes the form
$h_{\mu\nu} = Y^I h^I_{\mu\nu}$ and $h_{\alpha\beta} = Y^I
h^I_{\alpha\beta}$, where the indices $\mu, \nu, ...$ correspond
to the $AdS_5$ space, and the indices $\alpha, \beta$ correspond
to the $S_5$ space, and the $S^5$ harmonics $Y^I$ are defined in
the appendix.  The tensor perturbations are related to the scalars
$s^I$ by
\begin{eqnarray}
h^{I}_{\mu\nu} &=& -\frac{6k_I}{5}s^Ig_{\mu\nu} +
\frac{4R^2}{5(k_I + 1)}(5 \nabla_{\mu}\nabla_{\nu} -
g_{\mu\nu}\nabla_{\sigma}\nabla^{\sigma})s^I\nonumber\\
 &=&
- \frac{2k_I(k_I-1)}{(k_I+1)} g_{\mu\nu}s^I + \frac{4R^2}{(k_I +
1)}\nabla_{\mu}\nabla_{\nu}s^{I} , \nonumber\\ h^{I}_{\alpha\beta}
&=& 2k_I s^Ig_{\alpha\beta}\ . \label{pert}
\end{eqnarray}
Combining eqs.~(\ref{aofd}), (\ref{mink}), (\ref{norm}), and
(\ref{pert}) gives the metric perturbation.

The perturbation becomes much simpler in the Penrose limit.
Setting $t = x^+ + {x^-}/{R^2}$, $\psi =x^+ - {x^-}/{R^2}$, $\vec
u = {\vec w}/{R}$, $\vec v = {\vec y}/{R}$, and sending $R
\rightarrow \infty$, the general scalar perturbation~(\ref{mink})
becomes
\begin{equation}
\langle 0 |{\it T}\, \phi^I(t,\vec x)\, O^I(t',\hat x') \,| 0
\rangle \to \frac{{A(\Delta_I)}^{1/2} }{2^{\Delta_I}
\cos^{\Delta_I} [(1-i\epsilon)(x^+-t')] }\ .
\end{equation}
This leading behavior depends on the bulk position only through
$x^+$.  Also,
\begin{equation}
h_{MN} dx^M dx^N \to (h_{tt} + h_{\psi\psi}) dx^+ dx^+\ .
\end{equation}
For the non-BMN operator, the $SO(6)$ tensor ${\cal C}^I$ depends
only on the $SO(4)$ directions, and so $Y^I \to R^{-k_I} {\cal
C}^I(\vec y)$. Assembling all factors,
\begin{eqnarray}
h_{++} = h_{tt} + h_{\psi\psi} &=& \frac{4 R^2}{k_I+1} (k_I^2 +
\partial_t^2) s^I
\nonumber\\
&=& \frac{ R^{2-k_I}(k_I+1)\sqrt{k_I}}{N 2^{k_I/2}\cos^{k_I+2}
[(1-i\epsilon)(x^+-t')] } {\cal C}^I(\vec y)\ . \label{hpp}
\end{eqnarray}

Notice that the perturbation, for which $k_I = \tilde k_3$, has
the same $R^{2  - \tilde{k}_3}$ scaling as the gauge theory
amplitude~(\ref{finalmat}), since $J \propto R^2$.  Thus to obtain
a finite limit on either side we must scale the non-BMN operator
by $R^{k-2}$, as well as a factor of $N$ to offset the string
coupling.  The case $\tilde k_3=2$, whose limit is $1/N \times
{\rm finite}$, corresponds to a perturbation quadratic in the
$\vec y$, and so it is simply a time- and direction-dependent
modulation of the quadratic terms in the plane wave metric.
Higher $\tilde k_3$ correspond to nonlinearities which scale away
in the plane wave limit. Notice that the metric perturbation has
poles at infinitely many values of $x^+$.  This strange dependence
is due to the propagation of the bulk field from the $AdS_5$
boundary to the plane wave geodesic and will be discussed further
later.

This is the only metric perturbation that we will need for our
calculation. However, it is instructive to consider the form that
a metric perturbation derived from a string-like operator would
take. Consider, for example, the perturbation that would
correspond to $O^{I_1}$ at the boundary point $(z,\vec x) = (0,
\vec{0})$, which is the negative infinity of the global time.  In
this case,
\begin{equation}
\langle 0 | s^{I_1}(t,\vec x) \,| I_1 \rangle = \frac{2^{k_1/2}
(k_1+1)}{4 N \sqrt{k_1}} e^{-i k_1 t} (1 + u^2)^{-k_1/2}\ .
\end{equation}
The spherical harmonic is
\begin{equation}
Y^{I_1}(\psi,\vec v) = {k_1 \choose J} 2^{-J/2} e^{i J \psi}
(1-v^2)^{J/2} {\cal C}^{I_1}(\vec v)
\end{equation}
Now taking the Penrose limit with $\tilde k_1 = k_1 - J$ and $p_-
= -2J/R^2 $ fixed, the scalar perturbation becomes
\begin{equation}
\langle s^{I_1} \rangle Y^{I_1} \to \frac{|p_-|^{\tilde k_1 / 2}
\sqrt{J}}{4N\sqrt{\tilde k_1!}} e^{-i p_- x^- - i \tilde k_1 x^+ -
|p_-| (w^2 + y^2) / 4}\, {\cal C}^{I_1}(\vec y)\ . \label{pert1}
\end{equation}

Note the gaussian term which restricts the perturbation to the
plane $\vec w = \vec y = 0$.  Actually, to restrict to the
neighborhood of the limiting geodesic $x^- = 0$, we would need to
superpose different $J$-states to form a wavepacket in $x^-$; it
will not be necessary to do this explicitly.

This scalar perturbation scales as $R^{-2} \sim J^{-1}$.  There is
the explicit $\sqrt{J}$, an implicit $\sqrt{J}$ from the plane
wave behavior in $x^-$,\footnote{Roughly speaking, the range of
$x^-$ is $O(R^2)$ so we must divide by $R$ to obtain a normalized
state.} and a further $R^{-4}$ since $N^{-1} \propto \sqrt{G_{\rm
N}} R^{-4}$. The metric perturbation~(\ref{pert}) contains an
additional factor or $R^2 \sim J$ and so is finite in the Penrose
limit (there are terms in $h_{++}$ which scale as $J$, but these
cancel).  This is expected because this is the field corresponding
to a normalized string state. We could complete the calculation of
the three-point function directly as in LMRS, by evaluating the
trilinear supergravity amplitude (see also ref.~\cite{KKPR}).
However, this would only apply to chiral operators $|I_1\rangle$.
We will instead adopt a first-quantized description which applies
to general BMN operators.

Note the dependence of the perturbation~(\ref{pert1}) on $x_{-}$.
If we were to try to describe string propagation in this
supergravity background (which is not necessary to the
calculation), $p_-$ would no longer be a conserved quantity, and
in the light-cone description the three-string interaction would
change the string length.

\subsection{The string amplitude}

The three-point function~(\ref{finalmat}) is the amplitude for a
string in some initial state corresponding to $O^{I_1}$ to
propagate through the perturbed metric produced by $O^{I_3}$ and
come out as a string in the state produced by $O^{I_2}$.  This is
particularly simple in light-cone gauge, since the perturbation
just affects $g_{++}$:
\begin{equation}
g_{++} = -\frac{1}{2} z^2 +  h_{++}\ ,
\end{equation}
where $h_{++}$ is given by eq.~(\ref{hpp}) with $k \to \tilde k_3$
and $t' \to t_3$.  Also, $(z^1,\ldots,z^8) = (y^1,\ldots,y^4,
w^1,\ldots,w^4)$.  The action can be written down immediately,
\begin{equation}
S = \frac{1}{4\pi\alpha'} \int_0^{\pi\alpha' |p_-|} d\sigma
\left[\dot{{z}}^2 - {z}'^2 - {z}^2 +  h_{++}(x^+,\vec y) \right] +
\mbox{fermionic}\ .
\end{equation}
We will only consider bosonic excitations, so we do not need the
fermionic terms.
%  The metric perturbation that we are considering mixes
% with a five-form perturbation~\cite{LMRS} so there
% will indeed be a
% fermionic perturbation.  Elsewhere we will obtain this using
% supersymmetry~\cite{nmip}.

Expanding in the perturbation we have
\begin{equation}
\langle I_2 | O^{I_3}(x_3) | I_1\rangle = \frac{i}{4\pi\alpha'}
\int_{-\infty}^{\infty} dx^+ \int_0^{2\pi\alpha' |p_-|} d\sigma
\langle I_2 | h_{++}(x^+,\vec y) | I_1\rangle\ ,
\end{equation}
where the matrix element on the right-hand side is in the
light-cone one-string Hilbert space.  Expanding the fields in
creation and annihilation operators, the perturbation changes the
oscillator state of the incoming string.  Using
\begin{equation}
{z}^{i} = \sum_{n = -\infty}^{\infty} \frac{1
}{\sqrt{\omega_n|p_-|}}\left[ a_{n}^{i} e^{-i\omega_n x^+ + i n
\sigma/ \alpha'|p_-|} + a_{n}^{i\dagger}e^{i\omega_n x^+ - i
n\sigma/ \alpha'|p_-|} \right]\ ,
\end{equation}
the normalized initial and final states are
\begin{equation}
| I_a\rangle =
\frac{1}{\sqrt{\tilde{k}_a!}}\tilde{\mathcal{C}}^{I_1}_{i_1 ...
i_{\tilde{k}_a}} a_0^{i_1 \dagger} \cdots a_0^{i_{\tilde{k}_a}
\dagger}|0;p_-\rangle\ ,\quad a = 1,2\ .
\end{equation}
Because we are considering strings dual to chiral primaries in the
CFT, only zero modes are excited.  Thus in the perturbation we
need also keep only the zero modes of the $z$ fields.

In all,
\begin{equation}
\langle I_2 | h_{++}(x^+,\vec y) | I_1\rangle = \frac{(\tilde
k_3+1)\sqrt{\tilde k_3} R^{2-\tilde k_3} }{N 2^{\tilde
k_3/2}|p_-|^{\tilde k_3/2} \sqrt{\tilde k_1! \tilde k_2!} } \cdot
\frac{\langle{0}| \tilde C^{I_2}(\vec a_0) \tilde C^{I_3}(\vec a_0
e^{-i x^+}+ \vec a_0^\dagger e^{i x^+}) \tilde C^{I_1}(\vec
a_0^\dagger) | 0\rangle} {\cos^{\tilde k_3+2}
[(1-i\epsilon)(x^+-t_3)] }\ .
\end{equation}
The matrix element is
\begin{equation}
\frac{\tilde k_1 ! \tilde k_2! \tilde k_3 !}{\tilde\alpha_1 !
\tilde\alpha_2 !\tilde \alpha_3!} e^{-i (\tilde k_1 - \tilde k_2)
x^+} \langle
\mathcal{C}^{I_1}\mathcal{C}^{I_2}\mathcal{C}^{I_3}\rangle\ .
\end{equation}
The time integral is then
\begin{equation}
i \int_{-\infty}^\infty dx^+\, \frac{e^{-i (\tilde k_1 - \tilde
k_2) x^+}} {\cos^{\tilde k_3+2} [(1-i\epsilon)(x^+-t_3)] }
   = 2^{\tilde k_3 + 1} \frac{\tilde \alpha_1 ! \tilde
\alpha_2!} {(\tilde k_3 + 1)!}  e^{-i (\tilde k_1 - \tilde k_2)
x^+} \label{timeint}
% \equiv 2^{\tilde k_3 +1}
% I(\tilde{k}_2 - \tilde{k}_1,
% \tilde{k}_3 + 1)  \ .
\end{equation}
Collecting all factors, we have
\begin{equation}
\langle I_2 | O^{I_3}(t_3, \hat x_3) | I_1\rangle = \frac{ (|p_-|
R^2/2)^{1 - \tilde k_3/2} }{N} \frac{\sqrt{\tilde k_1! \tilde k_2!
\tilde k_3}}{\tilde \alpha_3!} \langle
\mathcal{C}^{I_1}\mathcal{C}^{I_2}\mathcal{C}^{I_3}\rangle
  e^{-i (\tilde k_1 - \tilde k_2) t_3}\ .  \label{finstring}
\end{equation}
This agrees with the field theory result~(\ref{finalmat}).

\sect{Discussion}

We should emphasize that the agreement between the CFT
amplitude~(\ref{finalmat}) and the string
amplitude~(\ref{finstring}) was a foregone conclusion.
LMRS~\cite{LMRS} have already shown that, due to supersymmetric
nonrenormalization, the free CFT and AdS three-point functions
match exactly.  We are simply repeating their calculation in a
different framework that is adapted to the plane wave limit and so
can be applies also arbitrary nonchiral BMN state, but for
external supergravity states our calculation must be equivalent to
theirs.\footnote{To be precise the supergravity description is
strictly valid only for $g N / J^2 \ll 1$, otherwise there is a
large invariant built from the string momentum and spacetime
curvature.  However, our string calculation shows that there are
no corrections involving this parameter.}

However, if we attempt to interpret this result in terms of a
holographic dual to the plane wave theory, things are not so
simple.  We would like to identify operators that are local in
time: a local operator at time $t'$ should correspond to a
perturbation of the plane wave spacetime that goes to $\delta(x^+
- t')$ as we approach the boundary. Even without being too
specific about what we mean by the boundary of the plane wave
spacetime, it is clear that this is not the case for the
perturbation~(\ref{hpp}) because the time dependence
\begin{equation}
\cos^{-k_I-2} [(1-i\epsilon)(x^+-t')] \label{timedep}
\end{equation}
factors out, and this does not become a delta function no matter
how we define and approach the boundary.

Of course, we began with a local operator in the AdS picture.  The
time dependence~(\ref{timedep}) arises from propagation from the
AdS boundary to the limiting geodesic.  In particular there are
poles when $x^+ - t'$ is an odd multiple of $\pi/2$.  This
represents the minimum propagation time from the AdS boundary to
the geodesic, plus multiple reflections from the AdS boundary.
This propagation is not part of the plane wave theory: the
factor~(\ref{timedep}) should be replaced with a delta-function to
obtain a local operator in the plane wave theory.  In other words,
the plane wave spacetime is essentially the tangent space to the
limiting geodesic, so that even its boundary is in an
infinitesimal neighborhood of the geodesic, and contributions
arising outside this neighborhood should be renormalized away.

Replacing the factor~(\ref{timedep}) by $\delta(x^+ - t)$ has a
puzzling effect on the amplitude: the factor
\begin{equation}
2^{\tilde k_3 + 1} \tilde \alpha_1 ! \tilde \alpha_2! / (\tilde
k_3 + 1)!
\end{equation}
 from the time
integral~(\ref{timeint}) is gone, and the result no longer agrees
with the free field theory calculation.  Further, noting that the
$\tilde\alpha_{1,2}$ depend jointly on the quantum numbers of all
three operators, agreement cannot be restored simply by
renormalization of the operators.

\setlength{\unitlength}{.4cm}
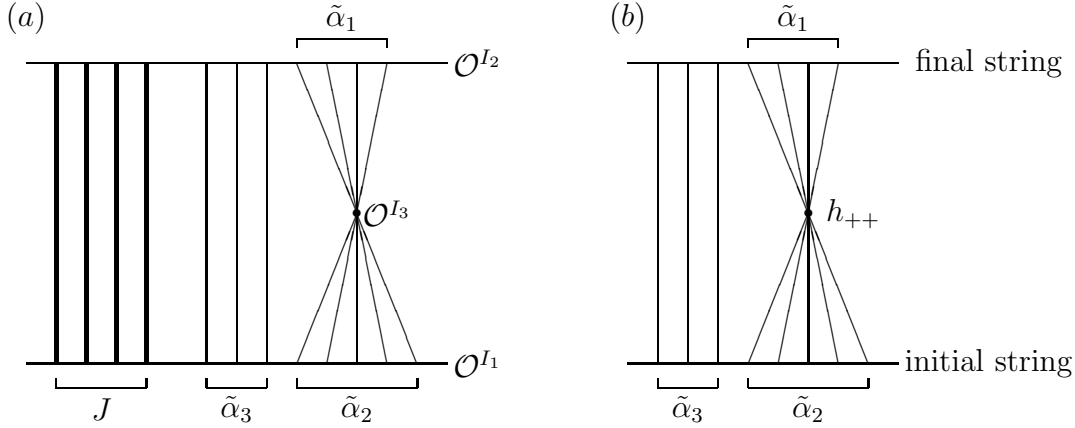
\begin{figure}
  \begin{picture}(36,16)
\put(0,2){\line(1,0){14}}
\put(15,2){\makebox(0,0){$\mathcal{O}^{I_1}$}}
\put(0,12){\line(1,0){14}} \put(0,13.5){\makebox(0,0){$(a)$}}
\put(15,12){\makebox(0,0){$\mathcal{O}^{I_2}$}}
\linethickness{.5mm} \put(1,2){\line(0,1){10}}
\put(2,2){\line(0,1){10}} \put(3,2){\line(0,1){10}}
\put(4,2){\line(0,1){10}} \thinlines \put(1,1.5){\line(0,-1){.3}}
\put(1,1.2){\line(1,0){3}} \put(4,1.2){\line(0,1){.3}}
\put(1,0){\makebox(3,1){${J}$}} \put(6,2){\line(0,1){10}}
\put(7,2){\line(0,1){10}} \put(8,2){\line(0,1){10}}
\put(6,1.5){\line(0,-1){.3}} \put(6,1.2){\line(1,0){2}}
\put(8,1.2){\line(0,1){.3}}
\put(6,0){\makebox(2,1){$\tilde{\alpha}_3$}}
\put(9,2){\line(2,5){2}} \put(10,2){\line(1,5){1}}
\put(11,2){\line(0,5){5}} \put(12,2){\line(-1,5){1}}
\put(13,2){\line(-2,5){2}} \put(9,1.5){\line(0,-1){.3}}
\put(9,1.2){\line(1,0){4}} \put(13,1.2){\line(0,1){.3}}
\put(9,0){\makebox(4,1){$\tilde{\alpha}_2$}}
\put(11,7){\circle*{.3}}
\put(12,7){\makebox(0,0){$\mathcal{O}^{I_3}$}}
\put(9,12){\line(2,-5){2}} \put(10,12){\line(1,-5){1}}
\put(11,12){\line(0,-5){5}} \put(12,12){\line(-1,-5){1}}
\put(9,12.5){\line(0,1){.3}} \put(9,12.8){\line(1,0){3}}
\put(12,12.8){\line(0,-1){.3}}
\put(9,13){\makebox(3,1){$\tilde{\alpha}_1$}}
\put(20,2){\line(1,0){9}} \put(32,2){\makebox(0,0){initial
string}} \put(20,12){\line(1,0){9}}
\put(20,13.5){\makebox(0,0){$(b)$}}
\put(32,12){\makebox(0,0){final string}}
\put(21,2){\line(0,1){10}} \put(22,2){\line(0,1){10}}
\put(23,2){\line(0,1){10}} \put(21,1.5){\line(0,-1){.3}}
\put(21,1.2){\line(1,0){2}} \put(23,1.2){\line(0,1){.3}}
\put(21,0){\makebox(2,1){$\tilde{\alpha}_3$}}
\put(24,2){\line(2,5){2}} \put(25,2){\line(1,5){1}}
\put(26,2){\line(0,5){5}} \put(27,2){\line(-1,5){1}}
\put(28,2){\line(-2,5){2}} \put(24,1.5){\line(0,-1){.3}}
\put(24,1.2){\line(1,0){4}} \put(28,1.2){\line(0,1){.3}}
\put(24,0){\makebox(4,1){$\tilde{\alpha}_2$}}
\put(26,7){\circle*{.3}} \put(27.5,7){\makebox(0,0){$h_{++}$}}
\put(24,12){\line(2,-5){2}} \put(25,12){\line(1,-5){1}}
\put(26,12){\line(0,-5){5}} \put(27,12){\line(-1,-5){1}}
\put(24,12.5){\line(0,1){.3}} \put(24,12.8){\line(1,0){3}}
\put(27,12.8){\line(0,-1){.3}}
\put(24,13){\makebox(3,1){$\tilde{\alpha}_1$}}
  \end{picture}
  \caption{The diagrams describing $(a)$ the CFT interaction and
  $(b)$ the world sheet interaction}
\end{figure}
It is interesting to examine in more detail the role that this
factor plays in the calculation.  Figure~1 shows the diagrams that
contribute to the CFT and world-sheet calculations.  We see that
they are the same diagram except that the CFT calculation has $Z$
contractions and counts only planar diagrams, whereas the world
sheet calculation counts both planar and non-planar graphs but has
no $Z$ contractions.  The similarities follow the usual plane-wave
intuition: the insertion of $\phi_i$ fields into the ${\rm Tr}\,
Z^J$ correspond to $a_0^{i\dagger}$ excitations of the string.

The following is then a summary of how the two sides of the
calculation come to be equal:
\begin{eqnarray}
&&\hspace{-10pt} \underbrace{\sqrt{(J + \tilde{k}_2)(J +
\tilde{k}_1)\tilde{k}_3}}_{{{\mbox{planar 3-point} \atop
\mbox{combinatorial}} \atop \mbox{factor}}} \times
\underbrace{\frac{J^{-\tilde{k}_3/2}\sqrt{\tilde{k}_1!\tilde{k}_2!}}
{\tilde{\alpha}_3!}}_{{\mbox{$Z$
field} \atop \mbox{contractions}}} \nonumber\\
&& \qquad= \underbrace{\frac{\tilde{k}_1!\tilde{k}_2!\tilde{k}_3!}
{\tilde{\alpha}_1!\tilde{\alpha}_2!\tilde{\alpha}_3!}}_{{
\mbox{oscillator}} \atop \mbox{contractions}}
 \times
\underbrace{\frac{\tilde{\alpha}_1!\tilde{\alpha}_2!}{(\tilde{k}_3
+ 1)!}}_{\mbox{time integral}} \times \underbrace{(p^+R^2)^{1 -
\tilde{k}_3/2} (\tilde{k}_3 +
1)\sqrt{\tilde{k}_3}}_{{\mbox{$h_{++}$} \atop
\mbox{normalization}}} \times
\underbrace{\frac{1}{\sqrt{\tilde{k}_1!\tilde{k}_2}!}}_{\mbox{
$O_{1,2}$} \atop \ \mbox{ norm.}}\ .\ \
\end{eqnarray}
If we focus on factors of $\tilde\alpha_i !$, which are the only
ones that cannot be absorbed into the normalizations of the
states, then the field theory has only a factor of
$\alpha_3!^{-1}$, which arises from the $\tilde \alpha_3 + J
\choose J$ permutations of the $Z$ fields among the $\tilde
\alpha_3 + J$ contractions between $O_1$ and $O_2$. The
world-sheet calculation has a factor of
$\tilde{\alpha}_1!\tilde{\alpha}_2!\tilde{\alpha}_3!$ from the
zero-mode contraction factor.  This factor accouunts for the
planar and non-planar contractions of the zero mode operators, and
there is no corresponding factor from the planar contractions in
the field theory.  Rather, the time integral brings in an
offsetting factor of $\tilde{\alpha}_1!\tilde{\alpha}_2!$.  For
holographic plane-wave operators defined as above, this factor is
absent.

Now let us try to understand the {\it disagreement} between the
free CFT calculation and the amplitudes obtained with the natural
holographic operators:

1) The most sweeping possibility is that there is a problem with
our basic premise that there is a holographic dual to the plane
wave string theory.  We see no reason to come to such a strong
conclusion limiting the generality of the holographic principle,
given the more conservative possibilities below.

2) It may be that the correct holographic observables are not the
local operators that we have attempted to define.  Indeed, there
are other contexts where a holographic dual exists but does not
have local operators~\cite{noloc}.  Further, it has been argued
that the geometry of the plane wave spacetime is such that the
natural observable is a two-dimensional $S$-matrix in the
$x^{\pm}$ directions~\cite{ppsm}, rather than the AdS-like
observables that we consider.  However, one of the authors of
ref.~\cite{ppsm} has informed us that there are difficulties with
this proposal.  Moreover, we note that the local operators that we
consider have a natural definition on the supergravity side, in
terms of metric perturbations proportional to delta functions in
$x^+$, so this suggests that they are naturally defined in the
dual theory as well; in the cases~\cite{noloc} without local
operators, the problem was already visible on the supergravity
side.  In a sense our amplitudes are the $S$-matrix: although the
perturbations are instantaneous in the light-cone time $x^+$, they
are propagating on null lines in spacetime --- this is the usual
pathology of the massless $S$-matrix in the light-cone in $1+1$
dimensions.

3) The above discussion of combinatorics suggests that the dual
theory might not be a planar theory.  This is intriguing, but
seems difficult to implement given that the construction of
strings as BMN states seems to depend in an essential way on the
planarity.

4) The simplest explanation is that the calculations do not match
because there are corrections from interactions on the CFT side.
These would be the analog of $gN$ corrections in the gauge theory,
but now in whatever is the smaller theory corresponding to the
plane wave spacetime.  The field theory calculations are valid for
small $gN$ and the string calculations are valid for large $gN$.
The absence of $gN$ corrections in other parts of the plane wave
duality is somewhat miraculous~\cite{GMR}, and may be due in part
to the large symmetry of the problem~\cite{SZ}.  There is no
strong reason to expect this to extend to the less BMN-like
observables that we consider. On the other hand, it is not clear
how to reconcile this with the nonrenormalization of the three
point function that was found in the full gauge
theory~\cite{LMRS}.

Spradlin and Volovich have independently considered the four-point
analog of our calculation (unpublished), comparing correlation
functions of two BMN operators and two non-BMN operators and
compared them with second-order perturbations of the world-sheet
theory.  Here the position-dependence is not fully determined by
conformal invariance. They also find a discrepancy between the
field theory and string theory results, suggesting the need for
$gN$ corrections.

Finally, we note again that metric perturbations with a
delta-function dependence in $x^+$ do seem to be natural
observables.  These would correspond to the operators of a quantum
mechanical theory, consistent with the conformal geometry of the
plane wave spacetime~\cite{ppholo}. Also, while we have considered
only operators with fixed $J$, it may be possible by appropriate
renormalization to define also local operators (at finite times)
with large $J$.

There are various interesting extensions of our calculation.  The
limiting procedure that we have considered gives operators that
depend on only four of the eight $z$-fields.  By taking instead
spherical harmonics on $S^3$ we can obtain operators with
arbitrary $z$-dependence. The extension to nonzero modes of the
string states is straightforward, and may give some additional
hint as to the origin of the discepancy that we have found.  It
would be interesting, but more challenging, to consider also
perturbations corresponding to excited string states.
Finally, we emphasize again the problem of determining the
nongravitational dual to string theory in the plane wave
spacetime.

\subsection*{Acknowledgements}

We would like to thank D. Berenstein, D. Gross, C. Keeler, M.
Sheikh-Jabbari, R. Roiban, M.~Sprad\-lin, L. Susskind, M. van
Raamsdonk, and A. Volovich for helpful discussions.  J.P. would
particularly like to thank R. Bousso for stimulating his interest
in plane wave spacetimes. This work was supported by National
Science Foundation grants PHY99-07949 and PHY00-98395.  The work of
N.M. was also supported by a National Defense Science and
Engineering Graduate Fellowship.

\appendix
\sect{Spherical harmonics}

Here we define the spherical harmonics that appeared in section~3.
We also repeat the calculation of the matrix element~(\ref{ccc})
in a slightly different way.  We continue the convention that the
reduced (3-sphere) objects are denoted with a tilde.  They are
normalized such that
\begin{equation}
\mathcal{C}^{I}_{i_1 \cdots i_k}x_{i_1}\cdots x_{i_k} = x^k
Y^{I}(x_1, ... ,x_6)
\end{equation}
and
\begin{equation}
\tilde{\mathcal{C}}^{I}_{i_1 \cdots i_{\tilde{k}}}x_{i_1}\cdots
x_{i_{\tilde{k}}} = x^{\tilde{k}}\tilde{Y}^{I}(x_1, ..., x_4)\ .
\end{equation}
With this normalization,
\begin{equation}
\frac{1}{\pi^3}\int_{S^5} Y^{I_1} Y^{I_2} =
\frac{\delta^{I_1I_2}}{2^{k-1}(k + 1)(k + 2)}\ ,
\end{equation}
\begin{equation}
\frac{1}{2\pi^2}\int_{S^3} \tilde{Y}^{I_1} \tilde{Y}^{I_2} =
\frac{\delta^{I_1I_2}}{2^{k}(k + 1)} .
\end{equation}
Defining
\begin{equation}
x_5 = x \cos \theta \cos \psi \ ,\quad x_6 = x \cos \theta \sin
\psi\ ,
\end{equation}
the spherical harmonics are related
\begin{eqnarray}
Y^{I_1} &=& 2^{-J/2}\sqrt{\frac{(J +
\tilde{k}_1)!}{J!\tilde{k}_1!}}e^{iJ\psi}\cos^J \!\theta\,
\sin^{\tilde{k}_1} \!\theta\, \tilde{Y}^{I_1}\ , \nonumber\\
Y^{I_2} &=& 2^{-J/2}\sqrt{\frac{(J +
\tilde{k}_2)!}{J!\tilde{k}_2!}}e^{-iJ\psi}\cos^J \!\theta\,
\sin^{\tilde{k}_2} \!\theta\, \tilde{Y}^{I_2}\ , \nonumber\\
Y^{I_3} &=& \sin^{\tilde{k}_3} \!\theta\, \tilde{Y}^{I_3}\ .
\end{eqnarray}
Then
\begin{eqnarray}
\langle \mathcal{C}^{I_1}\mathcal{C}^{I_2}\mathcal{C}^{I_3}
\rangle &=& (\Sigma/2 + 2)!2^{\Sigma/2 -
1}\frac{\alpha_1!\alpha_2!\alpha_3!}{k_1!k_2!k_3!}\times
\frac{1}{\pi^3}\int_{S_5} Y^{I_1}Y^{I_2}Y^{I_3}
\nonumber\\
&=& \frac{2^{\tilde{\Sigma}/2}(J + \tilde{\Sigma}/2 + 2)!(J +
\tilde{\alpha}_3)!\tilde{\alpha}_1!\tilde{\alpha}_2!}{\pi^2
J!\tilde{k}_3!\sqrt{\tilde{k}_1!\tilde{k}_2!(J + \tilde{k}_1)!(J +
\tilde{k}_2)!}}
\nonumber\\
&&\qquad \times \int_0^{\pi/2} \cos^{2J + 1} \theta
\sin^{\tilde{\Sigma} + 3} \theta d\theta \int_{S^3}
\tilde{Y}^{I_1}\tilde{Y}^{I_2}\tilde{Y}^{I_3}
\nonumber\\
&\to& \frac{J^{-\tilde{k}_3/2}\tilde{\alpha}_1\tilde{\alpha}_2}
{4\pi^2\tilde{k}_3!\sqrt{\tilde{k}_1!\tilde{k}_2!}} \times \int
e^{-y^2/2}(y^{\tilde{k}_1}\tilde{Y}^{I_1})
(y^{\tilde{k}_2}\tilde{Y}^{I_2}) (y^{\tilde{k}_3}\tilde{Y}^{I_3})
d^4 y \ .
\end{eqnarray}
Here $\Sigma = k_1 + k_2 + k_2$ and $\tilde\Sigma = \tilde k_1 +
\tilde k_2 + \tilde k_3$. The factor of $\cos^{2J + 1} \theta$
becomes a gaussian which in narrowly peaked on the geodesic
$\theta = 0$. Performing the $y$ integral, this reproduces the
right-hand side of eq.~(\ref{ccc}), in the large-$J$ limit. The
integral corresponds to a zero-dimensional field theory
calculation of the contractions of four scalar fields $y_i$. In
the world-sheet calculation of section~3, the zero mode
contractions have this same form.


\begin{thebibliography}{10}
\baselineskip=15pt

\bibitem{BMN}
D.~Berenstein, J.~M.~Maldacena and H.~Nastase, ``Strings in flat
space and pp waves from N = 4 super Yang Mills,'' JHEP {\bf 0204},
013 (2002) [arXiv:hep-th/0202021].
%%CITATION = HEP-TH 0202021;%%

\bibitem{BFHP}
M.~Blau, J.~Figueroa-O'Farrill, C.~Hull and G.~Papadopoulos,
``Penrose limits and maximal supersymmetry,'' Class.\ Quant.\
Grav.\  {\bf 19}, L87 (2002) [arXiv:hep-th/0201081].
%%CITATION = HEP-TH 0201081;%%

\bibitem{metplane}
R.~R.~Metsaev, ``Type IIB Green-Schwarz superstring in plane wave
Ramond-Ramond background,'' Nucl.\ Phys.\ B {\bf 625}, 70 (2002)
[arXiv:hep-th/0112044];

%%CITATION = HEP-TH 0112044;%%
R.~R.~Metsaev and A.~A.~Tseytlin,
``Exactly solvable model of superstring in plane wave Ramond-Ramond
background,'' Phys.\ Rev.\ D {\bf 65}, 126004 (2002)
[arXiv:hep-th/0202109].
%%CITATION = HEP-TH 0202109;%%

\bibitem{KPSS}
C.~Kristjansen, J.~Plefka, G.~W.~Semenoff and M.~Staudacher, ``A
new double-scaling limit of N = 4 super Yang-Mills theory and
PP-wave strings,'' Nucl.\ Phys.\ B {\bf 643}, 3 (2002)
[arXiv:hep-th/0205033].
%%CITATION = HEP-TH 0205033;%%

\bibitem{GMR}
D.~J.~Gross, A.~Mikhailov and R.~Roiban,
``Operators with large R charge in N = 4 Yang-Mills theory,''
Annals Phys.\  {\bf 301}, 31 (2002) [arXiv:hep-th/0205066].
%%CITATION = HEP-TH 0205066;%%

\bibitem{camb}
N.~R.~Constable, D.~Z.~Freedman, M.~Headrick, S.~Minwalla,
L.~Motl, A.~Postnikov and W.~Skiba,
``PP-wave string interactions from perturbative Yang-Mills theory,''
JHEP {\bf 0207}, 017 (2002) [arXiv:hep-th/0205089].
%%CITATION = HEP-TH 0205089;%%

\bibitem{KOS}
P.~Kraus, H.~Ooguri and S.~Shenker,
``Inside the horizon with AdS/CFT,''
arXiv:hep-th/0212277.
%%CITATION = HEP-TH 0212277;%%

\bibitem{GKPW}
S.~S.~Gubser, I.~R.~Klebanov and A.~M.~Polyakov, ``Gauge theory
correlators from non-critical string theory,'' Phys.\ Lett.\ B
{\bf 428}, 105 (1998) [arXiv:hep-th/9802109];
%%CITATION = HEP-TH 9802109;%%

E.~Witten, ``Anti-de Sitter space and holography,'' Adv.\ Theor.\
Math.\ Phys.\  {\bf 2}, 253 (1998) [arXiv:hep-th/9802150].
%%CITATION = HEP-TH 9802150;%%

\bibitem{ppholo}
S.~R.~Das, C.~Gomez and S.~J.~Rey, ``Penrose limit, spontaneous
symmetry breaking and holography in pp-wave background,'' Phys.\
Rev.\ D {\bf 66}, 046002 (2002) [arXiv:hep-th/0203164];
%%CITATION = HEP-TH 0203164;%%
%

E.~Kiritsis and B.~Pioline, ``Strings in homogeneous gravitational
waves and null holography,'' JHEP {\bf 0208}, 048 (2002)
[arXiv:hep-th/0204004];
%%CITATION = HEP-TH 0204004;%%
%

R.~G.~Leigh, K.~Okuyama and M.~Rozali, ``PP-waves and
holography,'' Phys.\ Rev.\ D {\bf 66}, 046004 (2002)
[arXiv:hep-th/0204026];
%%CITATION = HEP-TH 0204026;%%
%

D.~Berenstein and H.~Nastase, ``On lightcone string field theory
from super Yang-Mills and holography,'' arXiv:hep-th/0205048;
%%CITATION = HEP-TH 0205048;%%
%

D.~Marolf and S.~F.~Ross, ``Plane waves: To infinity and
beyond!,'' Class.\ Quant.\ Grav.\  {\bf 19}, 6289 (2002)
[arXiv:hep-th/0208197];
%%CITATION = HEP-TH 0208197;%%
%

G.~Siopsis, ``The Penrose limit of AdS x S space and holography,''
arXiv:hep-th/0212165;
%%CITATION = HEP-TH 0212165;%%
%

E.~T.~Akhmedov,
%``On the relations between correlation functions in SYM/pp--wave correspondence,''
arXiv:hep-th/0212297;
%%CITATION = HEP-TH 0212297;%%

T.~Yoneya, ``What is holography in the plane-wave limit of
AdS(5)/SYM(4) correspondence?,'' arXiv:hep-th/0304183.
%%CITATION = HEP-TH 0304183;%%

\bibitem{Vbit}
H.~Verlinde, ``Bits, matrices and 1/N,'' arXiv:hep-th/0206059.
%%CITATION = HEP-TH 0206059;%%

\bibitem{BKPSS}
N.~Beisert, C.~Kristjansen, J.~Plefka, G.~W.~Semenoff and
M.~Staudacher, ``BMN correlators and operator mixing in N = 4
super Yang-Mills theory,'' Nucl.\ Phys.\ B {\bf 650}, 125 (2003)
[arXiv:hep-th/0208178].
%%CITATION = HEP-TH 0208178;%%

\bibitem{3bmn}
M.~Spradlin and A.~Volovich, ``Superstring interactions in a
pp-wave background,'' Phys.\ Rev.\ D {\bf 66}, 086004 (2002)
[arXiv:hep-th/0204146];
%%CITATION = HEP-TH 0204146;%%

Y.~J.~Kiem, Y.~B.~Kim, S.~M.~Lee and J.~M.~Park, ``pp-wave /
Yang-Mills correspondence: An explicit check,'' Nucl.\ Phys.\ B
{\bf 642}, 389 (2002) [arXiv:hep-th/0205279].
%%CITATION = HEP-TH 0205279;%%

M.~X.~Huang, ``Three point functions of N = 4 super Yang Mills
from light cone string field theory in pp-wave,'' Phys.\ Lett.\ B
{\bf 542}, 255 (2002) [arXiv:hep-th/0205311];
%%CITATION = HEP-TH 0205311;%%

C.~S.~Chu, V.~V.~Khoze and G.~Travaglini, ``Three-point functions
in N = 4 Yang-Mills theory and pp-waves,'' JHEP {\bf 0206}, 011
(2002) [arXiv:hep-th/0206005];
%%CITATION = HEP-TH 0206005;%%

P.~Lee, S.~Moriyama and J.~W.~Park, ``Cubic interactions in
pp-wave light cone string field theory,'' Phys.\ Rev.\ D {\bf 66},
085021 (2002) [arXiv:hep-th/0206065].
%%CITATION = HEP-TH 0206065;%%

\bibitem{LMRS}
S.~M.~Lee, S.~Minwalla, M.~Rangamani and N.~Seiberg, ``Three-point
functions of chiral operators in D = 4, N = 4 SYM at large N,''
Adv.\ Theor.\ Math.\ Phys.\  {\bf 2}, 697 (1998)
[arXiv:hep-th/9806074].
%%CITATION = HEP-TH 9806074;%%

\bibitem{BKLT}
V.~Balasubramanian, P.~Kraus, A.~E.~Lawrence and S.~P.~Trivedi,
``Holographic probes of anti-de Sitter space-times,'' Phys.\ Rev.\
D {\bf 59}, 104021 (1999) [arXiv:hep-th/9808017].
%%CITATION = HEP-TH 9808017;%%

\bibitem{KKPR}
Y.~J.~Kiem, Y.~B.~Kim, J.~Park and C.~Ryou, ``Chiral primary cubic
interactions from pp-wave supergravity,'' JHEP {\bf 0301}, 026
(2003) [arXiv:hep-th/0211217].
%%CITATION = HEP-TH 0211217;%%

\bibitem{noloc}
A.~W.~Peet and J.~Polchinski, ``UV/IR relations in AdS dynamics,''
Phys.\ Rev.\ D {\bf 59}, 065011 (1999) [arXiv:hep-th/9809022];
%%CITATION = HEP-TH 9809022;%%

%
O.~Aharony and T.~Banks, ``Note on the quantum mechanics of M
theory,'' JHEP {\bf 9903}, 016 (1999) [arXiv:hep-th/9812237].
%%CITATION = HEP-TH 9812237;%%

\bibitem{ppsm}
D.~S.~Bak and M.~M.~Sheikh-Jabbari, ``Strong evidence in favor of
the existence of S-matrix for strings in plane waves,'' JHEP {\bf
0302}, 019 (2003) [arXiv:hep-th/0211073].
%%CITATION = HEP-TH 0211073;%%

\bibitem{SZ}
A.~Santambrogio and D.~Zanon, ``Exact anomalous dimensions of N =
4 Yang-Mills operators with large R charge,'' Phys.\ Lett.\ B {\bf
545}, 425 (2002) [arXiv:hep-th/0206079].
%%CITATION = HEP-TH 0206079;%%


%\bibitem{nmip}
%N.~Mann, work in progress.

\end{thebibliography}
\end{document}